\documentstyle[eqsecnum,aps,epsf] {revtex}

\begin{document}
\draft
\preprint{LAUR-94-****, gr-qc/94****}

\title{Geometrodynamic Quantization\\
       and Time Evolution in Quantum Gravity
       }

\author{Arkady Kheyfets\footnote{E-mail: kheyfets@odin.math.ncsu.edu}}
\address{Department of Mathematics,\break North 
   Carolina State University,\break Raleigh, NC 27695-8205}
\author{Warner A.~Miller\footnote{E-mail: wam@lanl.gov}}
\address{Theoretical Astrophysics Group (T-6, MS B288),\break 
   Theoretical Division,\break Los Alamos National Laboratory,\break
   Los Alamos, NM 87545}

\date{\today}

\maketitle

\begin{abstract}

 We advance here a new gravity quantization procedure that explicitly
utilizes York's analysis of the geometrodynamic degrees of freedom.
This geometrodynamic procedure of quantization is based on a
separation of the true dynamic variables from the embedding parameters
and a distinctly different treatment of these two kinds of variables.
While the dynamic variables are quantized following the standard
quantum mechanical and quantum field theoretic procedures, the
embedding parameters are determined by the ``classical'' constraint
equations in which the expectation values of the dynamic variables are
substituted in place of their classical values. This self-consistent
procedure of quantization leads to a linear Schr\"odinger equation
augmented by nonlinear ``classical'' constraints and supplies a
natural description of time evolution in quantum geometrodynamics. In
particular, the procedure sheds new light on the ``problems of time''
in quantum gravity.

\end{abstract}


\section{Introduction.} 
\label{I}

The task of describing time evolution in quantum gravity appears to
encounter numerous and seemingly insurmountable obstacles. The
difficulties are so pervasive that some researchers have voiced a
doubt whether the concept of time and time evolution can be adequately
introduced in quantum gravity.\cite{STC} The issue of time evolution
in quantum gravity has been reviewed by Kucha\v r\cite{Kuc92} who
formulated it as a set of ``problems of time.'' He concluded that the
problems of time have not been successfully resolved in either the
Dirac or the square-root ADM approaches to gravity quantization. In
this paper we review and analyze both these procedures of quantization
in light of a new approach to gravity quantization.\cite{KheMil94a}

The classical theory of gravity is a fully constrained theory.  It is
our thesis that this feature should not be mirrored in quantum
gravity, and our observation that the source of the ``time''
difficulties in both the Dirac and ADM quantization approaches arises
because the York split\cite{Yor72} of the dynamical variables from the
time parameterization variables occurs too late.  Here we propose to
introduce such a split before the dynamical theory is developed.
This, we suggest, will supply a more cogent quantization procedure as
well as bypass the ``problems of time.''

We provide in (Sec.~II) a brief review of the standard techniques of
canonical gravity quantization.  In Sec.~III we advance a new
procedure of gravity quantization (geometrodynamic quantization). This
procedure incorporates York's analysis of the geometrodynamic degrees
of freedom\cite{Yor72} and appears to avoid the difficulties of the
standard procedures of quantization. In Sec.~IV we provide two simple
examples to illustrate our geometrodynamic procedure of quantization;
namely, we quantize the axisymmetric Kasner and Taub cosmologies.
Sec.~V provides a brief discussion of the specific features of this
quantization approach, while Secs.~VI and VII provide a brief
description of the ``problems of time'' in standard canonical quantum
gravity and the solution of these problems within our new framework.

\section{Canonical Quantum Gravity.} 
\label{II} 

Both the Dirac and square-root Hamiltonian approaches to canonical
quantization make use of the 3+1 split of the spacetime geometry by
Arnowitt, Deser and Misner (hereafter referred to as ADM).\cite{ADM62}
According to the ADM prescription of classical general relativity one
considers a slicing of the spacetime by a family of spacelike
hypersurfaces labeled by a parameter $t$. This parameter can be
thought of as a time coordinate so that any slice is identified by the
relation $t = {\rm const}$. The remaining three spacetime coordinates
$x^i$ determine a coordinatization of each slice. The spacetime metric
(${}^{(4)}g_{\mu \nu}$) is parameterized by the shift ($N^i$), lapse
($N$) and the 3--metric of the slice ($g_{ik}$).
\begin{equation}
\label{1} 
\left(\matrix{{}^{(4)}g_{00} & {}^{(4)}g_{0k} \cr {}^{(4)}g_{i0} & {}^{(4)}g_{ik} 
\cr}\right) 
= \left(\matrix{(N_s N^s - N^2) & N_k \cr N_i & g_{ik} \cr}\right) . 
\end{equation}
The standard action of general relativity for the gravity field 
\begin{equation} 
\label{2} 
I = \int {\cal L} d^4x = \int{}^{(4)}R \sqrt{-{}^{(4)}g} d^4x, 
\end{equation} 
after its augmentation by appropriate boundary terms can be represented as 
\begin{equation} 
\label{3} 
I = \int\left[ \pi^{ij} {\partial g_{ij}\over \partial t} - N 
{\cal H}(\pi^{ij}, g_{ij}) - N_i {\cal H}^i(\pi^{ij}, g_{ij})\right] d^4x.
\end{equation} 
In this equation $\pi^{ij}$, ${\cal H}$, and ${\cal H}^i$ are given 
as follows\cite{MTW70}: 
\begin{equation} 
\label{4} 
\pi^{ij} = \left(\matrix{\hbox{\rm ``geometrodynamic} \cr 
\hbox{\rm field momentum''} \cr \hbox{\rm dynamically conjugate to} \cr 
\hbox{\rm the ``geometrodynamic} \cr 
\hbox{\rm field coordinate''} g_{ij} \cr}\right) = 
g^{1\over 2} (g^{ij}\, K - K^{ij}); 
\end{equation} 

\begin{equation} 
\label{5} 
{\cal H} = (\hbox{\rm ``super--Hamiltonian''}) = g^{-{1\over 2}} 
\left({\rm Tr}\Pi^2 - {1\over 2}\left({\rm Tr}\Pi\right)^2\right) - 
g^{1\over 2}R;\hbox{\ \ and} 
\end{equation} 

\begin{equation} 
\label{6} 
{\cal H}^i = (\hbox{\rm ``supermomentum''}) = -2 \pi^{ik}{}_{\vert k}; 
\end{equation}
where $K^{ij}$ is the extrinsic curvature tensor, $K=K^i_i$ is the
trace of the extrinsic curvature tensor, and $\Pi$ is the matrix of
$\pi^{ij}$. All the quantities in (\ref{4}) -- (\ref{6}), including
the covariant derivative in (\ref{6}) are related to the 3--geometry
of the slice.

Without any further refinement or redirection, the ADM approach treats
all six components of the 3--metric $g_{ik}$ as the gravitational
``field coordinates'' (with their conjugate momenta $\pi^{ik}$), while
the lapse $N$ and the shift $N^i$ are treated as Lagrange multipliers
(no conjugate momenta for them are present in the expression for
action).  The expression
\begin{equation} 
\label{7} 
{\cal H}_{ADM}(\pi^{ij}, g_{ij}, N, N_i) = N {\cal H}(\pi^{ij}, g_{ij}) + N_i 
{\cal H}^i(\pi^{ij}, g_{ij}) 
\end{equation} 
plays the part of Hamiltonian when variations of $\pi^{ij}$ and
$g_{ij}$ are considered. These variations produce the twelve Hamilton equations.
\begin{equation} 
\label{8} 
{\partial g_{ij}\over\partial t} = {\partial {\cal H}_{ADM}\over\partial 
\pi^{ij}} = 2 N g^{-{1\over 2}} 
\left( \pi_{ij} - {1\over 2}{\rm Tr}\Pi\right) + N_{i\vert j} + N_{j\vert i} 
\end{equation} 
\begin{equation} 
\label{9} 
{\partial\pi^{ij}\over\partial t} = - {\partial {\cal H}_{ADM}\over\partial 
g_{ij}}  
\end{equation}
We do not display this last equation in all detail as it is rather
lengthy and is not necessary for this paper. In addition, there are
three supermomentum constraints
\begin{equation} 
\label{10} 
{\cal H}^i(\pi^{ij}, g_{ij}) = 0
\end{equation} 
and the super--Hamiltonian constraint 
\begin{equation} 
\label{11} 
{\cal H}(\pi^{ij}, g_{ij}) = 0. 
\end{equation} 
These four constraint equations are obtained by variations of $N_i$
and $N$ in the ADM action, respectively. The equations (\ref{8}),
(\ref{9}), (\ref{9}), (\ref{10}) are not independent.  The interplay
between the Hamilton equations (\ref{8}), (\ref{9}) and the
constraints (\ref{10}), (\ref{11}) is rather involved. In particular,
if the constraints are satisfied on an initial slice then the Hamilton
equations guarantee that they are also satisfied on all spacelike
slices. On the other hand, if the constraints are satisfied on each
slice of all possible spacelike foliations of a given spacetime then
necessarily the Hamilton equations are satisfied. This last feature of
general relativity has led people to refer to gravitational dynamics
as a ``fully-constrained theory.'' While it is clear that this
property can be valuable in the classical theory, it is also equally
clear that it is implicitly based on the assumption of the uniqueness
of the spacetime 4--geometry \cite{Ger69}. This last assumption of
uniqueness becomes problematic in quantum geometrodynamics, and
consequently, the fully-constrained property of geometrodynamics also
becomes problematic.  This situation is, in our opinion, the root of
the notorious ``problems of time'' in quantum gravity.

A further investigation of the ADM picture of gravitational 
dynamics leads one to the interpretation of  the supermomentum constraints (\ref{10}) 
as expressing the 3--dimensional diffeomorphism invariance (the freedom of 
choice of coordinates on the slices) and the super--Hamiltonian constraint 
(\ref{11}), together with the natural identification \cite{MTW70} 
\begin{equation} 
\label{12} 
\pi^{ij} = {\delta S \over \delta g_{ij} } 
\end{equation} 
as the Hamilton--Jacobi equation of the theory \cite{MTW70,Per62}
\begin{equation} 
\label{13} 
{\cal H}\left({\delta S \over \delta g_{ij} }, g_{ij}\right) = 0.
\end{equation}  
Although this Hamilton--Jacobi equation involves all six components of the 
3--metric, the 3--diffeomorphic invariance as expressed by the supermomentum 
constraints, 
\begin{equation} 
\label{14} 
\left({\delta S \over \delta g_{ij} }\right)_{\vert j} = 0,
\end{equation} 
allows one to identify the Hamilton--Jacobi equation as an equation
that describes the evolution of the 3--geometry rather than of the
3--metric. Equation (\ref{13}) has numerous disturbing features. The
most disturbing of them is that, according to it, the
super--Hamiltonian ${\cal H}$ cannot be interpreted as the generator
of the time translation \cite{Mis57}. In addition, unlike the standard
situation in mechanics, the super--Hamiltonian ${\cal H}$
participating in the Hamilton--Jacobi equation (\ref{13}) does not
coincide with the ${\cal H}_{ADM}$ Hamiltonian (\ref{7}) which
generates the dynamic evolution via Hamilton equations (\ref{8}),
(\ref{9}).  This ${\cal H}_{ADM}$ Hamiltonian is related to the
Lagrangian ${\cal L}$ by the standard relation
\begin{equation} 
\label{15} 
{\cal H}_{ADM} = \pi^{ij} {\partial g_{ij}\over\partial t} - {\cal L} 
\end{equation}   
while, ordinarily, ${\cal H}$ is not. 

Dirac's procedure of canonical quantization utilizes the standard prescription  
\begin{equation} 
\label{16} 
\pi^{ij} \longrightarrow 
{\delta S \over \delta g_{ij} } \longrightarrow \widehat{\pi}_{ij} = 
{\hbar\over i} {\delta \over \delta g_{ij}} 
\end{equation} 
and leads to four functional differential equations based solely on the four 
constraint equations of the classical theory  
\begin{equation} 
\label{17} 
\widehat{{\cal H}}^i \left(\widehat{\pi}^{ij}, g_{ij} \right) \Psi = 0,
\end{equation} 
\begin{equation} 
\label{18} 
\widehat{{\cal H}} \left(\widehat{\pi}^{ij}, g_{ij} \right) \Psi = 0.
\end{equation} 
The three equations (\ref{17}) are often interpreted as a requirement
for the state functional to be a functional of the 3--geometry $\Psi =
\Psi [ {}^{(3)}{\cal G}]$ rather than of the 3--metric. Equation
(\ref{18}) is the Wheeler--DeWitt equation \cite{MTW70,DeW67} which is
considered to be a proper wave equation for quantum gravity.  This
equation reminds one more of a Klein--Gordon equation than a
Schr\"odinger equation. As we have already mentioned, the state
functional $\Psi$ in this equation, although originally introduced as
a functional of the 3--metric $\Psi = \Psi [g_{ik}]$, after imposing
on it the requirement of 3-dimensional diffeomorphic invariance, is
considered to be a functional of the underlying 3--geometry $\Psi =
\Psi [{}^{(3)}{\cal G}]$. The commutation relations are imposed on all
the components of the 3--metric, although proper care is taken to
ensure that they are diffeomorphically invariant.  The Wheeler-DeWitt
equation, just as its classical counterpart, the Hamilton--Jacobi
equation, does not admit an interpretation of the Hamiltonian as a
generator of time translations.

An attempt to remedy this situation via introducing, instead of the
super--Hamiltonian ${\cal H}$ a Hamiltonian that, in a sense, does
generate the time translation, has been undertaken in the ADM
square-root procedure of quantization\cite{Kuc92}. In this procedure
the set of six 3--metric parameters is split in two subsets
$\{\beta_1, \beta_2\}$, and $\{\alpha_1, \alpha_2, \alpha_3, \Omega\}$
of parameters. The first of them is treated as the set of true
gravitational degrees of freedom, and the second is considered as the
set of embedding variables. The classical super--momentum and
super--Hamiltonian constraints (\ref{10}), (\ref{11}) are then solved
on the classical level with respect to the momenta conjugate to the
embedding variables, which, after natural identifications (for the
sake of simplifications in notations we omit indices on $\beta$ and
$\alpha$ parameters; they can be recovered easily whenever it is
necessary) \cite{MTW70,DeW67}

\begin{equation} 
\label{19}    
\pi^\beta = {\delta S \over \delta\beta}, \qquad 
\pi^\alpha = {\delta S \over \delta\alpha}, \qquad 
\pi^\Omega = {\delta S \over \delta\Omega} 
\end{equation} 
leads to the following four equations:
\begin{equation} 
\label{20} 
{\delta S\over\delta\alpha} = - h_\alpha\left({\delta S \over\delta\beta}, \beta , 
\alpha , \Omega\right)
\end{equation} 

\begin{equation} 
\label{21} 
{\delta S\over\delta\Omega} = - h_\Omega\left({\delta S \over\delta\Omega}, \beta , 
\alpha , \Omega\right). 
\end{equation} 
One of the functions $h_\alpha$, $h_\Omega$ contains a square root operation in it 
(we assume from now on that it is $h_\Omega$), which is why it is called the 
square-root Hamiltonian. The standard prescription for quantization, 
\begin{equation} 
\label{22} 
{\delta S\over\delta\alpha} \longrightarrow i \hbar {\delta \over\delta\alpha}, 
\qquad 
{\delta S\over\delta\Omega} \longrightarrow i \hbar {\delta \over\delta\Omega}, 
\qquad 
{\delta S\over\delta\beta} \longrightarrow \widehat\pi^\beta = {\hbar\over i} 
{\delta \over\delta\beta}, 
\end{equation} 
leads to the four quantum equations, 
\begin{equation} 
\label{23} 
i \hbar {\delta \Psi \over\delta\alpha} = \widehat h_\alpha\left(\widehat\pi^\beta 
, \beta , \alpha , \Omega\right) \Psi, 
\end{equation} 
\begin{equation} 
\label{24} 
i \hbar {\delta \Psi\over\delta\Omega} = \widehat h_\Omega\left(\widehat\pi^\beta, 
\beta , \alpha , \Omega\right) \Psi,  
\end{equation} 
where $\Psi = \Psi\left[\beta , \alpha , \Omega\right]$. The choice of
parameters $\alpha$ and $\Omega$ is frequently made in such a way that
the three equations (\ref{23}) are interpreted as a requirement of the
diffeomorphism invariance, while equation (\ref{24}) is considered to
be the proper Schr\"odinger equation. In this case parameters $\alpha$
can be interpreted as coordinatization parameters while the function
$\Omega$ can be thought of as the many-fingered time parameter
\cite{BSW62,BSW63} with $h_\Omega$ treated as the Hamiltonian.

We wish to point out that the square-root Hamiltonian $h_\Omega$ in
the classical theory, just as it was the case with the
super-Hamiltonian ${\cal H}$, is not related to the Lagrangian ${\cal
L}$ as it would be in a standard dynamic theory (cf.~equation
(\ref{15})).

Both the ADM square-root and Dirac quantization procedures admit, in
principle, a time evolution prescription to be introduced in them,
although this should be done on different stages.  In Dirac's
quantization procedure the state functional $\Psi$ is a functional of
either the slice 3-metric or the slice 3-geometry, which can be
written in terms of $\beta$, $\alpha$, $\Omega$ functions as
\begin{equation} 
\label{25} 
\Psi = \Psi [{}^{(3)}g] = 
\Psi [\alpha ,\Omega , \beta ]\quad \hbox{\rm or} \quad 
\Psi = \Psi [{}^{(3)}{\cal G}] = \Psi [\Omega , \beta ],  
\end{equation} 
where $\alpha$, $\Omega$ and $\beta$ are treated on equal footing. In
particular, a computation of any quantity similar to an expectation
function, if it were possible, ordinarily would involve functional
integration over either all six functional parameters $\alpha$,
$\Omega$, $\beta$ or the three functional parameters $\beta$ and
$\Omega$. Along these lines we wish to remind the reader here that, in
Dirac's quantization approach, the commutation relations are imposed
on all the components of the 3-metric, and in quantum mechanics the
evaluation of expectations involves integration over all the variables
participating in the commutation relations. Such a calculation,
however, would ordinarily exclude a possibility for one to introduce a
functional time and a time evolution in the theory. A way to avoid
such an unpleasant situation is to split the set of variables in two
subsets, one of which ($\beta$) is to be considered as a set of true
functional arguments of the state functional (and to be integrated
over during computing the expectations), while the others ($\alpha$,
$\Omega$) are to be considered as a set of functional parameters (one
per slice) and to be interpreted as a representation of functional
time. This split should be introduced only after the solution for the
state functional is obtained and prior to the evaluation of
expectation values.  Such a procedure is thoroughly artificial and
amounts to treating the embedding variables as quantum ones in one
part of the theory yet as classical ones in another.

The square-root Hamiltonian quantization is based on the same
classical picture of the evolution in geometrodynamics as a change of
3--metric or 3--geometry from one spacelike slice to another and leads
one to a Hamilton--Jacobi equation. However, it introduces the split
of variables just before quantization. From this moment on it assumes
that there is only one function $\Omega$ per slice. The state
functional $\Psi$ when restricted to one slice becomes essentially a
functional of $\beta$ only, which can be properly written as
\begin{equation} 
\label{26} 
\Psi = \Psi[\alpha, \Omega ; \beta ]\quad \hbox{\rm or}\quad \Psi = 
\Psi[\Omega ; \beta ]. 
\end {equation} 
The computation of expectation functions (or any other momenta) now
assumes functional integration over the two parameters $\beta$ only,
with the result depending on $\Omega$ as a parameter (of course, it
will also depend on three other functional parameters $\alpha$, which
we call coordinatization parameters). This functional parameter ($\Omega$) plays
the role of the many-fingered time parameter. To summarize, in the square-root
Hamiltonian approach the switch from one picture of evolution to
another occurs at the final stage of formulating classical
geometrodynamics, and comes, just as in Dirac's approach, as an
afterthought aimed to meet the difficulties generated by the original
picture of evolution (as evolution of the 3--geometry) in the
quantized theory.  Such a switch does not cause any problems in the
classical theory, but comes at a price in quantum geometrodynamics by
generating numerous difficulties, including the notorious ``problems of
time'' (cf. Sec.~VI).

The equations of quantum gravity, be it equations (\ref{17}),
(\ref{18}) of Dirac's quantization or equations (\ref{23}), (\ref{24})
of the square-root Hamiltonian quantization, do not provide for a
natural split. The split is, to an extent, arbitrary. It reflects, for
any particular gravitational system, our understanding of the system's
dynamics. In a sense, it is similar to the situation described by
N.~Bohr regarding the split of any considered phenomenon into its
quantum and classical parts. Such a split depends not only on the
object that we are considering but, also, on the questions that we are
asking.

It appears to us that the equations of quantum gravity themselves do not 
imply the necessity of a picture of temporal evolution for
quantum gravitational systems. The assumption of a temporal evolution
should be added, at some level, so as to not contradict to the rest of
the theory.  The first step in this direction is to introduce a split
of metric variables into the true dynamic variables $\beta$ on one
hand, and the coordinatization parameters $\alpha$ and the
many-fingered time parameter $\Omega$ on the other hand. The variables
$\beta$ are the true quantum variables (calculation of expectation
values involves functional integration of $\Psi$ only over these
functions), while $\alpha$ and $\Omega$ are merely functional
parameters that allow one, after some manipulations, to introduce the
concept of a temporal evolution. The two different approaches
described in this section introduce the split of variables at
different stages.  This leads to difficulties of both the technical
and conceptual nature in quantum geometrodynamics.

In the next section we advance an alternative procedure of
quantization -- a procedure that, from the very beginning, is based on
a picture of geometrodynamic evolution induced by York's analysis of
the geometrodynamic degrees of freedom.  This alternative approach
does not involve a change in the paradigm of time evolution in
geometrodynamics.

\section{Geometrodynamic Quantization.}
\label{III}

We propose here an alternative approach to the quantization of
gravity.  Our approach is based on the post-ADM achievements made in
classical geometrodynamics.\cite{Whe88} In particular, we are
referring to York's solution of the initial--value problem and his
analysis of the gravitational degrees of freedom.\cite{Yor72,Yor73}
This development was initially motivated by Wheeler's semi--intuitive
remark that the 3--geometry of a spacelike hypersurface has encoded
within it the two gravitational degrees of freedom as well as its
temporal location within spacetime.  It is this notion that the
3--geometry is a carrier of information on time that has been referred
to as ``Wheeler's {\sl many--fingered time}.''\cite{BSW62,BSW63,MTW70}
It was J. York who first made this thesis precise.  He forwarded what
has now become almost the canonical split of the 3--geometry into its
underlying conformal equivalence class (its {\sl shape} representing
the two dynamic degrees of freedom of the gravitational field
coordinate per space point) and the conformal scale factor (its {\sl
scale} representing Wheeler's many-fingered time).  Only the conformal
3--geometry is truly dynamic in the sense that it can be specified
freely as the initial data. The scale factor is non-dynamic and
essentially specifies Wheeler's many-fingered time. The results of
York have demonstrated that the true dynamic part of the gravitational
field is not the 3-geometry but only its conformal part, and that the
proper configuration space or ``arena for geometrodynamics'' should be
the underlying conformal superspace (the space of all conformal
3-geometries) rather than Wheeler's superspace (the space of all
3-geometries).  The conformal scale factor and three other functional
parameters of the 3--metric (responsible for coordinate conditions)
thus become external parameters. In what follows we associate the true
dynamic variables $\beta$ with conformal 3-geometry, the many fingered
time variable $\Omega$ with the 3-geometry scale factor, and the
remaining three variables $\alpha$ of the 3--metric with a
coordinatization of a spacelike 3-surface. The many-fingered time
variable and coordinatization variables are assumed to be fixed
initially by some conditions, and then subsequently controlled by the
shift and lapse.

Our proposed procedure of gravity quantization is based, from the very
beginning, on York's analysis of gravitational degrees of freedom.  We
suggest that one should interpret geometrodynamics as an evolution of
the conformal 3--geometry $\beta$ in an external field determined by
the scale factor $\Omega$ and coordinatization variables
$\alpha$. Such an approach calls for a reformulation of
geometrodynamics on the classical level.  We start from the standard
Lagrangian ${\cal L}$ (written in terms of the 3--metric, shift and
lapse) and the associated action (with appropriate boundary terms, as
needed, to remove the second time derivatives terms) and we introduce
the momenta conjugate to the true dynamic variables
\begin{equation} 
\label{301} 
\pi_\beta = {\partial {\cal L}\over \partial\dot\beta}. 
\end{equation} 
We then use these $\pi_\beta$'s to form the geometrodynamic Hamiltonian 
${\cal H}_{DYN}$ of our approach, 
\begin{equation} 
\label{302} 
{\cal H}_{DYN} = \pi_\beta \dot\beta - {\cal L}. 
\end{equation} 
The new Hamiltonian ${\cal H}_{DYN}$ is distinctly different from 
${\cal H}_{ADM}$ and its arguments do not coincide with those of 
${\cal H}_{ADM}$, namely 
\begin{equation} 
\label{303} 
{\cal H}_{DYN} = {\cal H}_{DYN}(\Omega ,  
\alpha  ; \beta , \pi_\beta ). 
\end{equation}
 The variables preceding the semicolon are treated as describing an external field,
while the ones following the semicolon are the coordinates and
momenta of the gravitational true degrees of freedom, i.e. of the
conformal geometrodynamics. The variation of $\beta$ and $\pi_\beta$
leads to the equations of geometrodynamics, i.e. to two pairs of
Hamilton equations,
\begin{equation} 
\label{304} 
\dot\beta = {\partial {\cal H}_{DYN}\over\partial\pi_\beta}, 
\end{equation}  
\begin{equation} 
\label{305} 
\dot\pi_\beta = - {\partial {\cal H}_{DYN}\over\partial\beta}, 
\end{equation} 
and, subsequently, to the Hamilton--Jacobi equation 
\begin{equation} 
\label{306}  
{\delta S\over\delta t} = - {\cal H}_{DYN}\left(\Omega ,  
\alpha ; \beta , {\delta S\over\delta\beta}\right). 
\end{equation} 
Here $S$ is a functional of $\beta$ and, in addition, depends on the same 
parameters as ${\cal H}_{DYN}$, 
\begin{equation} 
\label{307} 
S = S\left[\Omega , 
\alpha ; \beta\right]. 
\end{equation} 

Neither the Hamilton equations (\ref{304}), (\ref{305}) nor the
Hamilton--Jacobi equation (\ref{306}) are capable of providing any
predictions as their solutions depend on the functional parameters
$\Omega$ and $\alpha$ which are not yet known.  One can complete the
system of equations by adding to the Hamilton equations, or to the
Hamilton--Jacobi equation, the standard constraint equations of
general relativity.  They should be satisfied when the solution 
for $\beta$, $\pi_\beta$ of
equations of conformal geometrodynamics (with appropriate initial 
data) is substituted in them (we use symbols $[\beta ]_s$, $[\pi_\beta ]_s$ 
for such a solution) 
\begin{equation} 
\label{308} 
 \matrix{{\cal H}^i\left(\Omega , \alpha , 
[\beta ]_s, [\pi_\beta ]_s\right) = 0 \cr 
  \cr 
{\cal H}\left(\Omega , \alpha ,  
[\beta ]_s, [\pi_\beta ]_s\right) = 0 \cr}
\end{equation} 
The resulting equations are equivalent to the standard equations of
classical geometrodynamics. It should be emphasized that these
constraint equations cannot be derived from variational principles in
our theory.

For the purpose of quantization, we start from our Hamilton--Jacobi
equation (\ref{306}), describing effectively what we refer to as {\sl
conformal geometrodynamics}. That is the evolution of the dynamic
variables corresponding to the conformal part of the 3--geometry,
where this evolution is parameterized by the ``external'' field
represented by (1) the scale parameter and (2) the coordinatization
variables.  Using the Hamilton--Jacobi equation (\ref{306}) we may
transition to the corresponding Schr\"odinger equation
\begin{equation} 
\label{309} 
i \hbar {\delta\Psi\over\delta t} = \widehat{\cal H}_{DYN}\left(\Omega ,  
\alpha ; \beta , \widehat\pi_\beta\right) \Psi 
\end{equation} 
where $\widehat\pi_\beta = {\hbar\over i} {\delta\over\delta\beta}$. 
The Schr\"odinger equation (\ref{309}) treats the scale
parameter and coordinatization functions as external classical fields 
and quantizes only the true dynamic variables, $\beta$. The state 
functional $\Psi$ in this equation is a functional of $\beta$ 
and also depends on the functional parameters $\Omega$ and $\alpha$. 
\begin{equation} 
\label{310} 
\Psi = \Psi\left[\Omega , 
\alpha ; \beta\right] 
\end{equation} 
This Schr\"odinger equation (with specific initial data) can be solved 
(cf., for instance the
example of the Bianchi~1A cosmological model in Sec.\ref{V}). The
resulting solution $\Psi_s$ of this Schr\"odinger equation is not capable of
providing any definite predictions as it depends on four functional
parameters $\Omega$, $\alpha$ which remain at this stage
undetermined. All expectations, such as, for instance the expectation
values of $\beta$
\begin{equation} 
\label{311} 
<\beta >_s = \langle\Psi_s\vert\beta\vert\Psi_s\rangle = 
\int\Psi^*_s \beta \Psi_s \, {\cal D}\beta 
\end{equation}
or of $\widehat\pi_\beta$ 
\begin{equation} 
\label{312} 
<\pi_\beta >_s = \langle\Psi_s\vert\widehat\pi_\beta\vert\Psi_s\rangle = 
\int\Psi^*_s \widehat\pi_\beta \Psi_s \, {\cal D}\beta 
\end{equation} 
also depend on these functional parameters.  To specify these
functions we resort to the constraint equations. As in case of
classical geometrodynamics the constraints should be imposed on the
solution of the initial-value problem of conformal geometrodynamics,
and in this way, determine the unique values of $\Omega$ and
$\alpha$. It is possible that there are several ways to couple the
constraints to the quantization of the true dynamic variables,
$\beta$.  We propose here that the four constraints be imposed only on
the expectation values of the conformal dynamics
\begin{equation} 
\label{313} 
\matrix{{\cal H}^i\left(\Omega , \alpha , 
<\beta >_s, <\pi_\beta >_s\right) = 0 \cr 
  \cr 
{\cal H}\left(\Omega , \alpha ,  
<\beta >_s, <\pi_\beta >_s\right) = 0 \cr}
\end{equation}  
i.e. only on measurable quantities.  In so doing, we explicitly
avoid the interpretational conundrums associated with the problems of
time, and we form a ``classical'' gravitational clock driven by the
quantized geometrodynamic system -- i.e. {\sl quantum-driven
many-fingered time}.

This procedure of quantization utilizes explicitly the correct
treatment of the geometrodynamic degrees of freedom and introduces a
meaningful time parameterization utilizing the shift and lapse. The
super-momentum and super-Hamiltonian constraints are not satisfied as
strictly as in the Dirac and ADM square-root approaches. Our point of
view is that the question of slicing independence of evolution is not
a well posed question in quantum gravity and should be recovered only
in the classical limit. It is clear that this weakening of the
constraints, so that they hold only on the expectation values over the
solution of the initial-value problem of the Schr\"odinger equation,
makes the theory less restrictive and enlarges the set of possible
solutions.

The goal of this paper is limited to a clarification of our basic
thesis regarding the quantization of gravity --- quantum
geometrodynamics.\cite{KheMil94a} We believe that the best way to
achieve this goal is to consider a couple of simple illustrative
examples.

\section{Quantum Geometrodynamics: The Kasner and Taub Cosmologies}
\label{IV}

\subsection{Quantum Geometrodynamics of the Bianchi 1A Cosmology.} 
\label{IV-A}

The Bianchi 1A cosmological model is commonly referred to as the
axisymmetric Kasner model \cite{RS75}. Its metric is determined by two
parameters, the scale factor $\Omega$ and the anisotropy parameter
$\beta$ (we choose $N^i = 0$ and $N =1$ values of shift and lapse for
this example)
\begin{equation}
\label{401}
ds^2 = - dt^2 + {\rm e}^{-2\Omega} \left( {\rm e}^{2\beta} dx^2 + 
{\rm e}^{2\beta} dy^2 + {\rm e}^{-4\beta} dz^2 \right).
\end{equation}
As this cosmology is homogeneous the two functions $\Omega$ and $\beta$ are 
the functions of the time parameter $t$ only. The scalar 4--curvature can 
be expressed in terms of these two functions to yield the Hilbert action and, 
after subtracting  the boundary term, the cosmological action,
\begin{equation}
\label{402}
I_C = I_H + {3V\over 8\pi}\dot\Omega {\rm e}^{-3\Omega}\vert_{t_0}^{t_f} = 
{3V\over 8\pi}\int\limits_{t_0}^{t_f} \left( \dot\beta^2 - \dot\Omega^2 
\right) {\rm e}^{-3\Omega} dt,
\end{equation}
where $V = \int\int\int dx dy dz$ is the spatial volume element.

We treat the scale factor $\Omega (t)$ as the many-fingered time parameter and 
the anisotropy $\beta (t)$ as the dynamic degree of freedom. The momentum 
conjugate to $\beta$ is 
\begin{equation}
\label{403}
p_\beta = {\partial L \over \partial\dot\beta} = {3V\over 4\pi} 
{\rm e}^{-3\Omega} \dot\beta.
\end{equation} 
(as it is usually the case with homogeneous cosmologies, we are working here 
with the momentum $p_\beta$ rather than with the density $\pi_\beta$). 
The Hamiltonian of the system in our approach can be expressed in terms of 
the momentum conjugate to $\beta$ and the Lagrangian.
\begin{eqnarray}
\label{404}
H_{DYN} & = & p_\beta \dot\beta - L \nonumber \\ 
  & = & {4\pi \over 3V} {\rm e}^{3\Omega} p_\beta^2 - {3V \over 8\pi} \left( 
        {4\pi \over 3V}\right)^2 {\rm e}^{3\Omega} p_\beta^2 + {3V\over 8\pi} 
        \dot\Omega^2 {\rm e}^{-3\Omega} \nonumber \\ 
  & = & {2\pi \over 3V} {\rm e}^{3\Omega} p_\beta^2 + {3V\over 8\pi} 
\dot\Omega^2 {\rm e}^{-3\Omega}. 
\end{eqnarray}
In the classical theory this Hamiltonian can be used to produce either one
pair of Hamilton equations or the equivalent Hamilton--Jacobi
equation.  In any case, the dynamics picture derived in this way is
incomplete. To complete it we impose the super-Hamiltonian
constraint.
\begin{equation}
\label{405}
p_\beta^2 = \left(  {3V \over 4\pi}\right)^2 {\rm e}^{-6\Omega} \dot\Omega^2.
\end{equation}

Using the Hamilton--Jacobi equation,
\begin{equation}
\label{406}
{\partial S\over \partial t} = - H_{DYN}\left( {\partial S\over \partial\beta}, 
\Omega (t), \dot\Omega (t) \right),
\end{equation}
together with the expression (\ref{404}) for the Hamiltonian $H$ and the standard 
quantization prescription we obtain the Schr\"odinger equation for 
the axisymmetric Kasner model.
\begin{equation}
\label{407}
i\hbar {\partial\Psi\over\partial t} = -{2\pi\hbar^2 \over 3V} 
{\rm e}^{3\Omega} {\partial^2 \Psi \over \partial\beta^2} + 
{3V\over 8\pi} \dot\Omega^2 {\rm e}^{-3\Omega} \Psi .
\end{equation}
The constant $\hbar$ in this equation should be understood as the square of 
Planck's length scale, rather than the standard Planck constant. We wish to 
stress here that the scale factor $\Omega$ in the Schr\"odinger equation is 
so far an unknown function of time. This means that the equation does not 
describe completely the quantum dynamics of the axisymmetric Kasner model.  
To complete the dynamics picture we follow our prescription and impose, in 
addition to equation (\ref{407}), the super-Hamiltonian constraint.
\begin{equation}
\label{408}
<p_\beta>_s^2 = \left( {4\pi \over 3V}\right)^2 {\rm e}^{-6\Omega} \dot\Omega^2.
\end{equation}
Here $<p_\beta>_s$ is the expectation value of the momentum $\widehat p_\beta = 
{\hbar\over i} {\partial \over \partial\beta}$ 
\begin{equation}
\label{409}
<p_\beta>_s = \langle\Psi_s\vert\widehat p_\beta\vert\Psi_s\rangle = 
\int\limits_{-\infty}^{\infty} \Psi_s^*(\beta , t) \widehat p_\beta \Psi_s 
(\beta , t) d\beta 
\end{equation}
where $\Psi_s$ is the solution of the Schr\"odinger equation with 
specified initial data.
The system of equations (\ref{407}), (\ref{408}) provide us with a complete
quantum dynamic picture of the axisymmetric Kasner model and, when
augmented by appropriate initial and boundary conditions, can be
solved analytically.

Before discussing the initial value conditions we will find the general 
solution of the Schr\"odinger equation considering the scale factor $\Omega$ 
as a function of time generating an external potential. For this we separate 
variables,
\begin{equation}
\label{410}
\Psi (\beta , t) = \phi (\beta ) T(t).
\end{equation}
After substituting (\ref{410}) in the Schr\"odinger equation (\ref{407}) we obtain, 
\begin{equation}
\label{411}
i\hbar \phi \dot T = -{2\pi\hbar^2\over 3V} {\rm e}^{3\Omega} T {\phi}'' + 
{3 V\over 8\pi}\dot\Omega^2 {\rm e}^{-3\Omega} T \phi,
\end{equation}
where the prime means differentiation with respect to $\beta$. Rewriting it as 
\begin{equation}
\label{412}
{2\pi\hbar^2\over 3V} {\phi''\over\phi} = -i\hbar {\rm e}^{-3\Omega} 
{\dot T\over T} + {3V\over 8\pi} {\rm e}^{-6\Omega}\dot\Omega^2 = -\lambda,
\end{equation}
where $\lambda$ is the constant of separation, we obtain the equations for 
$\phi (\beta )$ and $T(t)$. 
\begin{equation}
\label{413}
\phi'' + {3V\over 2\pi\hbar^2} \lambda \phi = 0
\end{equation}
\begin{equation}
\label{414}
{\dot T\over T} = -{i\over\hbar} {\rm e}^{3\Omega} \left( {3V\over 8\pi} 
{\rm e}^{-6\Omega} \dot\Omega^2 + \lambda \right) 
\end{equation}
Equation (\ref{413}) admits only positive eigenvalues for $\lambda$.
Introducing the notation ${3V\lambda\over 2\pi} = k^2$ we can write
the solutions $\phi_k(\beta )$, $T_k(t)$ for $k \in (-\infty , \infty
)$.
\begin{eqnarray}
\label{415}
\phi_k(\beta) & = & A_k {\rm e}^{{i\over\hbar}k\beta} \nonumber \\ 
T_k(t) &  =  & B_k \exp\left\{ {-{i\over\hbar}\int_{t_0}^t \left( {2\pi\over 
               3V} k^2 +{3V\over 8\pi} {\rm e}^{-6\Omega}\dot\Omega^2 \right) 
               {\rm e}^{3\Omega} dt}\right\}
\end{eqnarray}
Using the superposition of these solutions we come up with the general 
solution of the Schr\"odinger equation (\ref{407}).
\begin{equation}
\label{416}
\Psi (\beta, t) = \int\limits_{-\infty}^{\infty} A_k 
{\rm e}^{{i\over\hbar}k\beta} 
\exp\left\{ {-{i\over\hbar}\int_{t_0}^t \left( {2\pi\over 
3V} k^2 +{3V\over 8\pi} {\rm e}^{-6\Omega}\dot\Omega^2 \right) 
{\rm e}^{3\Omega} dt}\right\} dk 
\end{equation}

To specify a particular problem one has to furnish appropriate initial data.
\begin{equation}
\label{417}
\Psi (\beta , t)\vert_{t_0} = \Psi (\beta , t_0) = 
\int\limits_{-\infty}^{\infty} A_k {\rm e}^{{i\over\hbar}k\beta} dk 
\end{equation}
It can be done either by specifying a function $\Psi(\beta,t_0)$ and then 
recovering $A_k$ from the equation
\begin{equation}
\label{418}
\Psi (\beta , t_0) = 
\int\limits_{-\infty}^{\infty} A_k {\rm e}^{{i\over\hbar}k\beta} dk,
\end{equation}
using Fourier transforms, or by assigning $A_k$ as a function of $k$, 
depending on the type of the problem to be formulated. In this section we 
consider the simplest example comparable with the quantum 
mechanics of a particle, namely a wave packet. To describe a Gaussian wave 
packet centered initially at the value $k_0$ of $k$ (we will describe the 
meaning of $k_0$ later) we assign 
\begin{equation}
\label{419}
A_k = C {\rm e}^{-a(k - k_0)^2},
\end{equation}
where the constant $a$ effectively determines the initial width of the wave 
packet in momenta and  $C$ is the normalization constant. This leads to the 
following 
expression for the initial values of the wave function:
\begin{equation}
\label{420}
\Psi (\beta ,t_0) = C \int\limits_{-\infty}^{\infty} {\rm e}^{-a(k - k_0)^2} 
{\rm e}^{{i\over\hbar}k\beta} dk = C \sqrt{{\pi\over a}} 
{\rm e}^{{i\over\hbar}\beta k_0} {\rm e}^{-{\beta^2\over 4a\hbar^2}}.
\end{equation}
The value of the normalization constant $C$ is determined by the condition 
\begin{equation}
\label{421}
\langle\Psi\vert\Psi\rangle = C^2 {\pi\over a} \int\limits_{-\infty}^{\infty} 
{\rm e}^{-{\beta^2\over 2 a \hbar^2}} d\beta = C^2\hbar \pi^{{3\over 2}} 
\sqrt{{2\over a}} = 1 
\end{equation}
which leads to the value of $C^2$ 
\begin{equation}
\label{422}
C^2 = {\sqrt{a} \over \hbar\pi^{{3\over 2}}\sqrt{2}}.
\end{equation}
Using expression (\ref{419}) for $A_k$ and introducing notations for f and g, 
\begin{eqnarray}
\label{423}
f = f(t) = {2\pi\over 3V}\int\limits_{t_0}^t {\rm e}^{3\Omega} dt, \nonumber \\
g = g(t) = {3V\over 8\pi} \int\limits_{t_0}^t \dot\Omega^2 
                    {\rm e}^{-3\Omega} dt,
\end{eqnarray}
we can write down the solution $\Psi_s(\beta, t)$ for the wave packet.
\begin{equation}
\label{424}
\Psi_s(\beta, t) = C {\rm e}^{-{i\over\hbar}g} \int\limits_{-\infty}^{\infty} 
{\rm e}^{-a(k -k_0)^2} {\rm e}^{{i\over\hbar}\beta k} {\rm e}^{{i\over\hbar} 
f k^2} dk.
\end{equation}
After a simple transformation this expression can be rewritten in the form,
\begin{equation}
\label{425}
\Psi_s(\beta ,t) = C \exp\left\{{i\over\hbar} \left[ (\beta - k_0 f) k_0 - 
g\right]\right\} \int\limits_{-\infty}^{\infty} {\rm e}^{-ak^2} 
{\rm e}^{-{i\over\hbar}fk^2} {\rm e}^{{i\over\hbar}(\beta - 2k_0f)k} dk.
\end{equation}
The integral on the right hand side of (\ref{425}) can be evaluated. The final 
expression for the solution describing a Gaussian wave packet may be 
written in the following form which will prove to be convenient for
future calculations: 
\begin{eqnarray}
\label{426}
\Psi_s(\beta ,t)=C \sqrt{\pi} \left( a^2 + {f^2\over\hbar^2} 
\right)^{-{1\over 4}} 
\cdot \exp\left\{ -{a\over 4\left( a^2 + {f^2\over\hbar^2}\right)}  
{(\beta - 2k_0f)^2\over\hbar^2} \right\} \times \nonumber \\ 
\exp\left\{ {i\over\hbar} (\beta - k_0f) k_0 \right\} 
\cdot \exp\left\{ i{{f\over\hbar} \over 4\left( a^2 + {f^2\over\hbar^2}\right)} 
{(\beta - 2k_0f)^2\over\hbar^2} \right\} \cdot \exp\left\{ -{i\over\hbar}g - 
i \theta \right\};
\end{eqnarray}
where,
\begin{equation} 
\label{427}  
\cos ({2\theta}) = a/\sqrt{a^2 + f^2/\hbar^2}, \qquad 
\sin ({2\theta}) = (f/\hbar)/\sqrt{a^2 + f^2/\hbar^2} \nonumber
\end{equation}
Although expression (\ref{425}) looks quite involved the last three
exponential factors are phase factors and do not complicate the
determination of the expectation values of the observables.

It is clear that this solution of the Schr\"odinger equation
describing the wave packet cannot provide any definite predictions as
it contains the two functions of time $f(t)$ and $g(t)$ which are
themselves related to the as yet undetermined scale factor $\Omega$.
To find $\Omega(t)$ we need to (1) compute the expectation
$<p_\beta>_s$ of the momentum $\widehat p_\beta = {\hbar\over i}
{\partial \over\partial\beta}$, (2) substitute this expectation value
into the constraint (\ref{408}) and (3) solve the resulting equation
with respect to $\Omega$. We start from computing $<p_\beta>_s$.
\begin{equation}
\label{428}
<p_\beta>_s = \langle\Psi_s\vert\widehat p_\beta \vert\Psi_s\rangle = \\ 
C^2 \pi \left( a^2 + {f^2\over\hbar^2}\right)^{-{1\over 2}} k_0 
\int\limits_{-\infty}^{\infty} \exp \left\{ -{a\over 2\left( a^2 + 
{f^2\over\hbar^2}\right)} 
{(\beta - 2k_0f)^2\over\hbar^2} \right\} d\beta = k_0. 
\end{equation}
In other words the expectation value of the momentum $<p_\beta >_s$ does not 
change with time. It is determined by the $k$--center of the packet at 
$t=t_0$. Substitution of this result in (\ref{31}) yields 
\begin{equation}
\label{429}
k_0^2 = \left( {3V \over 4\pi}\right)^2 {\rm e}^{-6\Omega} \dot\Omega^2.
\end{equation}
This equation and the classical equations are identical. Therefore,
we need not describe it in detail. We only wish to point out once more
that after the solution of this equation is substituted in (\ref{426})
the geometrodynamic problem (\ref{407}), (\ref{408}) for the wave packet
(\ref{419}) is solved completely. To summarize, the many-fingered time
of quantum geometrodynamics in case of a Gaussian wave packet of
axisymmetric Kasner spacetimes coincides with its classical
counterpart if the expectation value of the momentum of the packet is
identified with the (conserved) value of the momentum of the classical
solution.

The expectation value for the anisotropy parameter $\beta$, where
$\beta$ is the only quantum dynamic variable in this model, is given
by:
\begin{equation}
\label{430}
<\beta >_s = \langle\Psi_s\vert\beta\vert\Psi_s\rangle = \\ 
C^2\pi \left(a^2 + {f^2\over\hbar^2}\right)^{-{1\over 2}} 
\int\limits_{-\infty}^{\infty} \beta \exp\left\{ -{a\over 2\left( a^2 + 
{f^2\over\hbar^2}\right)} {(\beta - 2k_0f)^2\over \hbar^2} \right\} d\beta 
= 2k_0f(t).  
\end{equation}
Thus ``the center'' of the wave packet evolves as the classical Kasner
universe determined by the momentum value equal to $k_0$ would evolve.
The spread of the wave packet with time is the variance in $\beta$.
\begin{eqnarray}
\label{431}
<\left(\beta\  - <\beta >\right)^2>_s = \nonumber \\ 
C^2\pi \left(a^2 + {f^2\over\hbar^2}\right)^{-{1\over 2}} 
\int\limits_{-\infty}^{\infty} (\beta - 2k_0f)^2 \exp\left\{ -{a\over 2\left( 
a^2 + {f^2\over\hbar^2}\right)} {(\beta - 2k_0f)^2\over \hbar^2} \right\} d\beta = 
{\hbar^2 a^2 + f^2 \over a}
\end{eqnarray}
It is obvious from (\ref{431}) that the spread of the packet increases
with time.  The result is similar to that of the quantum mechanics of
a free particle; after all the Bianchi~I cosmology is the
free--particle analogue of quantum cosmology.

\subsection{Quantum Geometrodynamics of the Taub Cosmology.} 
\label{IV-B}

We present within this section a second application of our approach to
quantum gravity. In addition to illustrating our theory on a
relatively simple model, we show that very concept of time emerges by
imposing the principle of general covariance as weakly as
possible.\cite{KheMil94a} In particular, the Hamiltonian constraint
is imposed as an expectation-value equation over the true dynamic
degree of freedom of the Taub cosmology -- a representation of the
underlying anisotropy of the 3-space.  In this way the concept of time
appears to be inextricably intertwined and woven to the initial
conditions as well as to the quantum dynamics over the space of all
conformal 3-geometries.  This quantum geometrodynamic approach will
ordinarily lead to quantitatively different predictions than either
the Dirac or ADM quantizations, and in addition, our approach appears
to avoid the interpretational conundrums associated with the
``problems of time.''\cite{Kuc92} So without further ado, in this
subsection we apply our quantum geometrodynamic approach to the Taub
cosmology \cite{RS75} and numerically solve the coupled Schr\"
odinger and expectation-value constraint equations.

The Taub cosmology is an axisymmetric homogeneous cosmology
parameterized by a scale factor $\Omega(t)$, and an anisotropy
parameter $\beta(t)$.  The line element may be expressed as
\begin{equation}
\label{t1}
ds^2=-dt^2+a_{\circ}^2 e^{2\Omega}\left(e^{2\beta}\right)_{ij}\sigma^i\sigma^j,
\end{equation}
where $(\beta)=diag(\beta,\beta,-2\beta)$, and the one forms as,
$\sigma^1=\cos\psi d\theta+\sin\psi d\phi$,
$\sigma^2=\sin\psi d\theta-\cos\psi d\phi$, and
$\sigma^3=d\psi+\cos\theta d\phi$.  The scalar 4-curvature is
expressed in terms of $\Omega$ and $\beta$ and yields the 
action,
\begin{equation}
\label{t2}
I_c = \frac{3\pi a_{\circ}^3}{4}\int\{(\dot{\beta}^2-\dot{\Omega}^2)
                                        -\frac{1}{6} ^{(3)}R\}
                                   e^{3\Omega} dt.
\end{equation}
Here $^{(3)}R=\frac{e^{-2\beta}}{2a_{\circ}^2e^{2\Omega}}(4-e^{-6\beta})$
represents the scalar 3-curvature.

	We treat here the scale factor $\Omega(t)$ as the many-fingered time 
parameter \cite{BSW62} and the anisotropy $\beta(t)$ as the dynamic
degree of freedom.\cite{Yor72}  The momentum conjugate to $\beta$ is obtained
from from the Lagrangian, ${\cal L}$.
\begin{equation}
\label{t3}
p_\beta = \frac{\partial {\cal L}}{\partial \dot{\beta}} 
        = \frac{3}{2}\pi a_{\circ}^3 e^{3\Omega} \dot{\beta}
        = m\dot{\beta}
\end{equation}
The dynamical Hamiltonian for this cosmology can be expressed in terms of
this momentum and the Lagrangian,
\begin{equation}
\label{t4}
{\cal H}_{DYN} = p_{\beta}\dot{\beta} - {\cal L} 
        = \frac{1}{2m}p_{\beta}^2 + 
           \underbrace{\frac{m}{2}\left(\dot{\Omega}^2-
                       \frac{1}{6}^{(3)}R\right)}_{V}.
\end{equation}
In the classical theory ${\cal H}_{DYN}$ can be used to construct either the 
Hamilton-Jacobi equation or the two Hamilton equations; however, to 
complete the dynamics we must in addition
impose the super-Hamiltonian constraint,
\begin{equation}
\label{t5}
p_{\beta}^2 = m^2\left(\dot{\Omega}^2+\frac{1}{6}^{(3)}R\right).
\end{equation}

	Using this dynamical Hamiltonian (Eq.~(\ref{t4})), the corresponding
Hamilton-Jacobi equation and the standard quantization prescription we
obtain the Schr\" odinger equation for the Taub cosmology.
\begin{equation}
\label{t6}
-i\hbar \frac{\partial \Psi (\beta,t)}{\partial t} =
-\frac{\hbar^2}{2m}\frac{\partial^2\Psi (\beta,t)}{\partial \beta^2} +
V\Psi (\beta,t)
\end{equation}
The scale factor $\Omega$ in $V$ and $m$ in this equation should be
treated as an unknown function of time, $t$.  To complete the
quantization we impose, in addition, Eq.~(\ref{t5}) as an expectation-value
equation over $\beta$ (where $<\bullet>_s=\int\Psi_s^*\bullet\Psi_s d\beta$ and 
$\Psi_s$ is the solution of of (\ref{t6}) augmented by appropriate initial data, 
cf.~Eq.~(\ref{t8})).
\begin{equation}
\label{t7}
\dot{\Omega}^2 = \left(\frac{<p_{\beta}>_s}{m}\right)^2 +
                 \frac{1}{6}<^{(3)}R>_s
\end{equation}
This system of equations Eq.~(\ref{t6}) and Eq.~(\ref{t7}) provide us with a
complete quantum-dynamic picture of the axisymmetric Taub model, and
when augmented by the appropriate boundary conditions can be solved
numerically.

	We display in Figs.~1-2 a solution of the these two equations 
for an initially Gaussian wave packet.
\begin{equation}
\label{t8}
\Psi_0=C e^{i(\beta-\beta_{\circ})p_{\circ}/\hbar}
         e^{-(\beta-\beta_{\circ})^2/\Delta\beta^2}
\end{equation}
with $C^2=\sqrt{2/\pi}/\Delta\beta$, $p_{\circ}=-100$, $\hbar=1$,
$\beta_{\circ}=0$, $\Delta\beta=0.1$ and $\Omega(t=0)=1$.  The solution 
was obtained using a 2nd-order split operator unitary integrator which 
preserves the norm exactly (up to round off errors).\cite{FeiFle83}  

\begin{figure}
\centerline{\epsfxsize=6.0truein\epsfbox{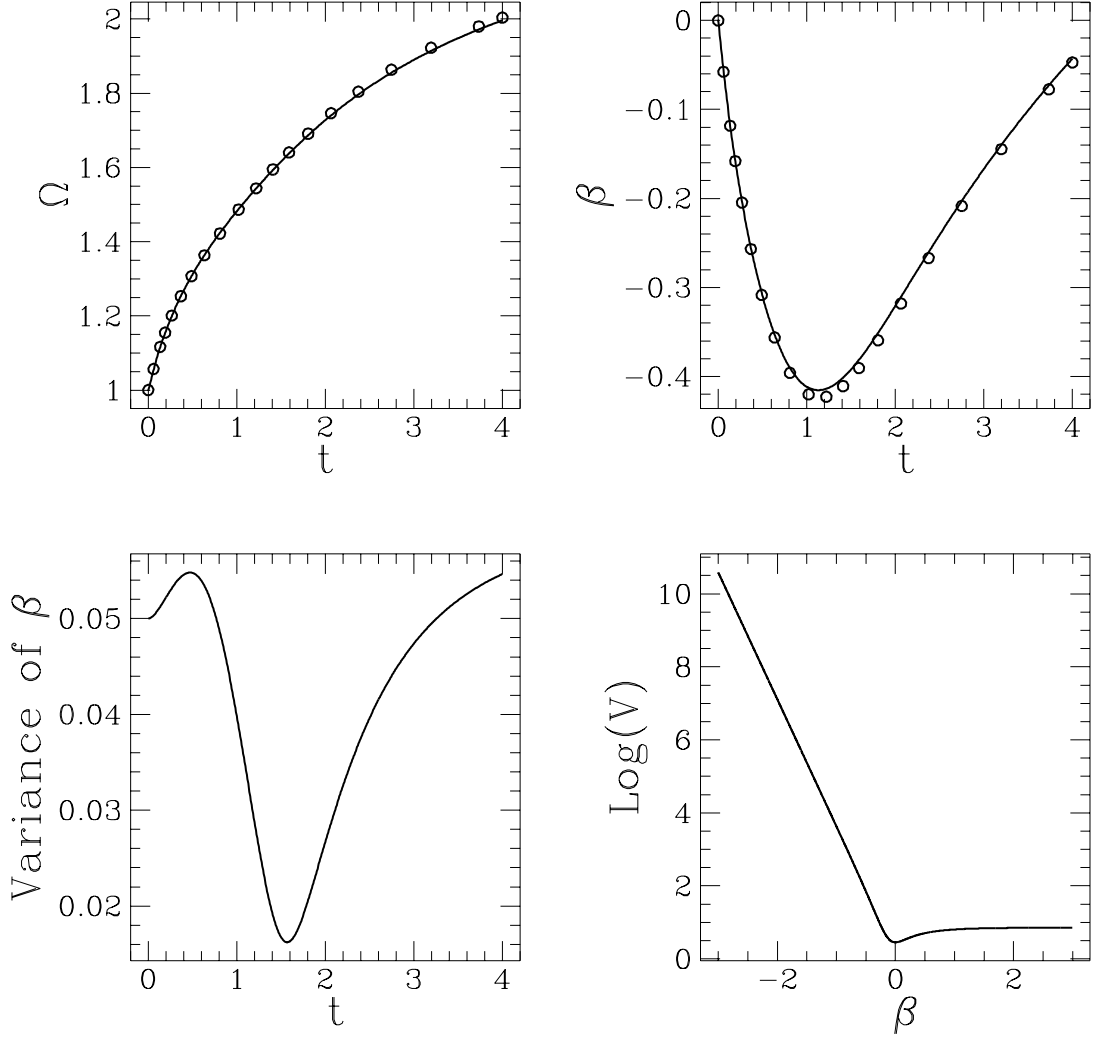}}
\label{Fig1}
\caption{
{\em The quantum geometrodynamics of the Taub cosmology.}  The solid
lines in each of the four graphs represent the quantum solution, while
the small circular dots in the upper two graphs represent the
classical trajectory.  The scale factor's ($\Omega$) dependence on
time ($t$) is shown in the upper left, while the expectation of the
anisotropy ($<\beta>$) is displayed in the upper right.  The lower
left shows the variance in $\Psi_s$ throughout its evolution, and the
lower right graph is a snapshot of the potential ($V$) at the end of
the simulation.}
\end{figure}

\begin{figure}
\centerline{\epsfxsize=6.1truein\epsfbox{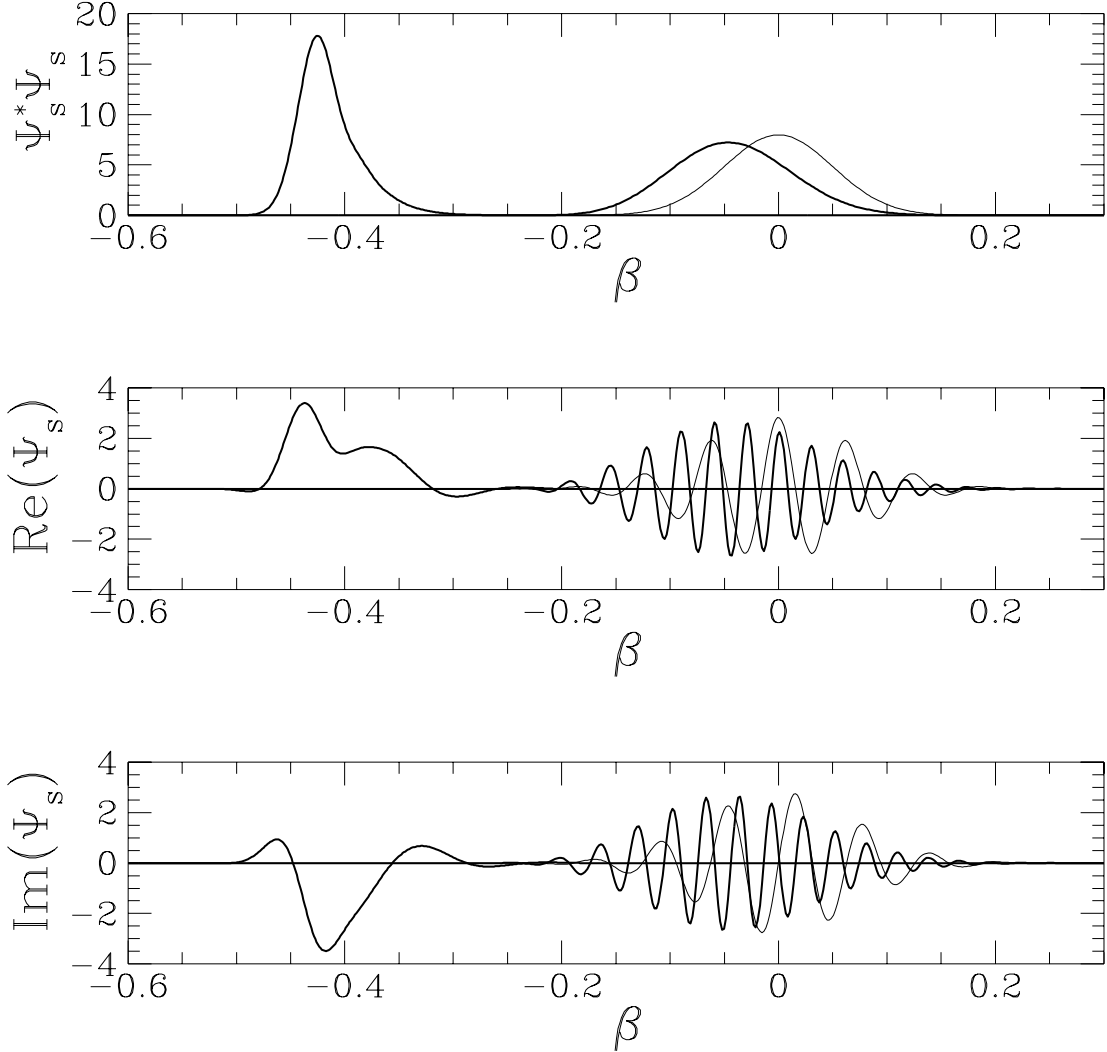}}
\label{Fig2}
\caption{
{\em The Taub wave function.}  A snapshot of the wave function at
three distinct times throughout the evolution is represented on each
of the three graphs. The first row plots $\Psi^*_s\Psi_s$, while the
second and third are the real and imaginary parts of $\Psi_s$,
respectively.  The right-most wave function (plotted in a slightly
narrower line width) shows the initial ($t=0$) gaussian wave function.
The left-most and higher-peaked curve represents the wave function near the
bounce at $t=1$, while the remaining curve displays the right-going wave function 
well after the bounce at $t=4$.  This reflected wave function is spreading 
in time and therefore appears, and will continue to appear, lower in amplitude than
the initial wave function.}
\end{figure}

	We have introduced in this section an application of our
approach to the quantization of gravity (this approach need not be
restricted to quantum cosmology).  When the Taub cosmology was
quantized our theory generated a quantum picture based on a post-ADM
treatment\cite{Yor72} of the gravitational field dynamics and appears
to be free from the conceptual difficulties related with time usually
associated with both the Dirac and ADM quantization procedures.  Only
the dynamical part, the underlying anisotropy ($\beta$) has been
quantized.  The Hamiltonian participating in the quantization (${\cal
H}_{DYN}$) is not a square-root Hamiltonian.  This absence of a
square-root Hamiltonian is a generic feature of our quantum
geometrodynamics procedure.

	We have demonstrated here using the Taub cosmology that the concept
of time may very well be inextricably intertwined and woven to the initial
conditions as well as to the quantum dynamics over the space of all 
conformal 3-geometries.

\section{The Characteristic Features of the Quantum Geometrodynamic Procedure.}
\label{V}

The quantum dynamical picture described in the two previous sections has
two features that are not encountered typically in most of the more
common quantum dynamical schemes. They should be kept in mind in order
to avoid errors and misinterpretations in applying our procedure.

The first such feature is related to the structure of the Hamiltonian.
Formally, the Hamiltonian appears as a Hamiltonian of a system placed
in external time-dependent field, at least when the Schr\"odinger
equation is analyzed. However, the external field is determined via
the constraints by the quantum state of the system.  Ordinarily, the
Hamiltonian depends on the initial data.  This feature is not unique
for our approach as it is encountered elsewhere in standard quantum
mechanics of Hartree--Fock systems.

The second feature is a split of the system parameters in two groups:
(1) the truly dynamic variables (to be quantized); and (2) the
descriptors of a ``natural observer'' associated to the system, driven
as they are by the quantum geometrodynamics. Such a split is
introduced prior to quantization, and is crucial to our theory.  This
feature can be clearly observed in both the general description of the
geometrodynamic quantization procedure (Sec.~III) and in the examples
of Sec.~IV. Essentially, this split is a part of our solution of the
time problems. Most of the difficulties related to the problems of time
are caused by an attempt to enforce general covariance at the quantum
level. Such an attempt for gravitational systems is bound to fail. It
leads to the questions that are not well defined unless they are
referred to a background spacetime. Standard attempts to introduce a
background spacetime tend to use a fixed spacetime with properties
generally not related to the properties of the system itself, which is
commonly considered as an unsatisfactory feature (we quite agree with
this conclusion). Our procedure, on the other hand, can be considered
as a universal prescription for defining a unique background spacetime
structure driven by, and back reacting on, the quantum system. All
questions concerning covariance should be referred to this spacetime
structure, and this spacetime structure is ordinarily not
classical. It cannot be considered as a result of a 3--geometry
evolution (the classical evolution equations do not take a part in our
procedure). In particular, we recover the classical evolution only
through an appropriately--peaked wave function and only through
constructive interference over conformal superspace.

The split of the metric variables into quantized geometrodynamic
variables and ``classical'' or non-quantized variables amounts to a
representation of a quantum gravitational system as two interacting
subsystems. One of the subsystems is quantum while the other one is
classical. Evolving the many-fingered time variable and
coordinatization variables by the shift and lapse is just another
aspect of such a subdivision of the quantum gravitational system. From
this point of view the functional evolution problem and the multiple
choice problem are simply ill posed. Different time evolutions or
different quantum theories referred to in the description of these
problems are essentially created by introducing different physical
systems, or by introducing different models of what is believed to be
the same system. In other words, quantum geometrodynamics yields a
finer resolution of different systems than classical
geometrodynamics. The reason that this split occurs is related to the
key features of gravitational systems. In particular, the observer in
gravitational systems is a part of the system, and the measurement
process (coupling of a measuring device with the quantum subsystem)
cannot be switched on and off at will.  In quantum cosmology a
particular split of variables into quantum and classical essentially
reflects our view concerning the nature of time in the Universe.
Whether this view is correct or not can be decided only on the basis
of observations.

We wish to emphasize here that the square-root Hamiltonian
quantization procedure essentially works in a way similar to our
geometrodynamic quantization procedure. It starts from the split of
variables in embedding parameters (they are to become the functional
parameters of the state functional, essentially classical) and dynamic
variables, that are to be quantized. A computation of the expectation
values of observables involves functional integration over the dynamic
variables while the embedding parameters are not integrated
over. However the approach does not contain a meaningful prescription
of time parameterization. An attempt to reintroduce such a prescription
using shift and lapse parameterization of time evolution (cf. Sec.~VI)
leads to the theory demonstrating the same features as our
procedure. The spectral analysis problem characteristic to this
approach is a consequence of an attempt of one to transfer to quantum
gravity the fully-constrained property of classical geometrodynamics.
Our procedure forfeits this feature and instead introduces a system of
equations that is not overdetermined, and for which the concept of
being fully constrained looses its meaning. Nevertheless, this
property is recovered in the classical limit only. As a result, the
geometrodynamic approach does not introduce the square-root
Hamiltonian.

Not every assignment of dynamical variables leads to a consistent
quantum dynamical picture. For example, a bad choice of the dynamical
component of the 3--geometry might lead to a non-elliptic differential
operator on the right hand side of Schr\"odinger equation.
Furthermore, under such conditions not every shape of the initial wave
packet will agree with the constraints. In any case, only an
appropriate choice of the set of dynamic variables will lead to a
reasonable quantum mechanical picture. The properties of the ADM
procedure together with the results of York indicate that there is
at least one such reasonable choice. Is there more than one possible 
quantum dynamical picture for the same system? From our point of 
view, the existing theory is ill equipped to ask this question. The 
notion of equivalence of quantum gravitational systems is simply not 
developed. Extreme care should be exercised when one 
attempts to discuss this question naively.

\section{The Problems of Time.} 
\label{VI}

The issues related to time evolution in gravity have been summarized
recently by Kucha\v r.\cite{Kuc92} He calls them the ``problems of
time'' in quantum gravity.

The first of the time problems is formulated as a dependence of the
final state $\Psi_{\Sigma_{FIN}}$ at a given $\Psi_{\Sigma_{IN}}$ on
the foliation that connects the spacelike hypersurfaces $\Sigma_{IN}$
and $\Sigma_{FIN}$. This problem is called the problem of functional
evolution.

The second problem is observed by Kucha\v r in the context of the
square-root Hamiltonian quantization. It stems from the fact that the
split of the metric variables into the true dynamic variables $\beta$ as opposed
to the coordinatization parameters $\alpha$ and the slicing parameter
$\Omega$ is not unique. Different splits might lead to different and
possibly inequivalent quantum theories. Kucha\v r calls this the
multiple-choice problem.

The third difficulty is the so called Hilbert space problem and is
particular to the Dirac quantization procedure. In the square-root
Hamiltonian approach, the Schr\"odinger equation automatically
determines an inner product and a possibility to form, at given values
of $\alpha$ and $\Omega$ meaningful observables, for which expectation
values can be evaluated in a way similar to that of standard quantum
mechanics (modulo some technical difficulties caused by the functional
nature of variables). Nothing similar is granted by the
Wheeler--DeWitt equation of the Dirac approach. The Wheeler--DeWitt
equation is a second order functional differential equation for which
the problem of supplying the space of its solution with the
structure of a Hilbert space encounters numerous difficulties. Roughly
speaking, the equation does not admit a one--Universe interpretation.

We wish to add to these three problems one more (also described by
Kucha\v r), namely the spectral-analysis problem that shows up if the
square-root Hamiltonian is involved. It is well known that, if the
operator expression under the square root is not positively definite,
the spectral analysis procedure yields, as the result, a Hamiltonian
that is not self--adjoint. This implies that the Schr\"odinger
equation with such a Hamiltonian does not produce unitary evolution.

The functional-evolution problem and the multiple-choice problem are,
essentially, two aspects of the same problem. Both these
problems in quantum gravity are caused by (1) the ADM treatment of the
whole 3--geometry as dynamic, and (2) the fully-constrained feature of
the dynamics of the gravitational field. In both procedures of
quantization (Dirac and ADM) these two requirements are expressed via
the demand for the classical constraints to yield operator equations
that should be satisfied for ``any slicing'' (or, as they frequently
say, for any parameterization). The concept of ``any parameterization'',
however, is essentially classical, which is especially clear when it
is thought of as a slicing. The concept of slicing works quite well in
the classical theory where the 4--geometry of spacetime is unique, but
does not have any analogue in the quantum theory unless we introduce
one and give it an appropriate meaning. We believe that both problems
emerge as a result of different assumptions concerning the split of a
gravitational system into its classical and quantum parts which
interact with each other. Different splits ordinarily provide
physically different models of the gravitational system and cannot be
compared to each other in the absence of a common spacetime. In other
words, the concept of equivalent systems is absent in the theory and
should be developed prior to posing these questions.

Although there are many different ways to specify coordinatization
parameters and slicing parameters, not all of them are meaningful in
the sense that the resulting time can be related to measurements. A
standard procedure to introduce a meaningful time in general
relativity uses the lapse and shift functions. In this approach the
coordinatization and slicing conditions are given arbitrarily only at
the initial slice.  Subsequently, both the coordinatization and
slicing conditions are propagated via specifying shift $N^i$ and lapse
$N$ functions. In order to facilitate an explanation of the
consequences of such a description for the square-root Hamiltonian
quantization procedure, it is useful to rewrite the action principle
and Hamilton equations in terms of the $\beta$, $\alpha$, and $\Omega$
parameters.

The expression (\ref{3}) for the action can be written symbolically in the 
form 
\begin{equation} 
\label{27} 
I = \int\left[ \pi^\beta \dot\beta + \pi^\alpha \dot\alpha + \pi^\Omega 
\dot\Omega  - N 
{\cal H}(\pi^\beta , \pi^\alpha , \pi^\Omega , \beta , \alpha , \Omega ) - 
N_i {\cal H}^i(\pi^\beta , \pi^\alpha , \pi^\Omega , \beta , \alpha , 
\Omega )\right] d^4x, 
\end{equation} 
where 
\begin{equation} 
\label{28} 
\pi^\beta = {\partial {\cal L}\over \partial\dot\beta}, \qquad 
\pi^\alpha = {\partial {\cal L}\over \partial\dot\alpha}, \qquad 
\pi^\Omega = {\partial {\cal L}\over \partial\dot\Omega}, 
\end{equation} 
are  the momenta conjugate to $\beta$, $\alpha$, $\Omega$ respectively. 
Expression (\ref{7}) takes the form 
\begin{equation} 
\label{29} 
{\cal H}_{ADM}(\pi^\beta , \pi^\alpha , \pi^\Omega , \beta, \alpha, \Omega , 
N, N_i) = N {\cal H}(\pi^\beta , \pi^\alpha , \pi^\Omega , \beta , \alpha , 
\Omega ) + N_i {\cal H}^i(\pi^\beta , \pi^\alpha , \pi^\Omega , \beta , 
\alpha , \Omega ). 
\end{equation} 
The Hamilton equations (\ref{8}), (\ref{9}) become 
\begin{equation} 
\label{30} 
\dot\beta = {\partial {\cal H}_{ADM}\over\partial\pi^\beta}, 
\end{equation} 
\begin{equation} 
\label{31} 
\dot\alpha = {\partial {\cal H}_{ADM}\over\partial\pi^\alpha}, 
\end{equation} 
\begin{equation} 
\label{32} 
\dot\Omega = {\partial {\cal H}_{ADM}\over\partial\pi^\Omega}, 
\end{equation} 
and 
\begin{equation}
\label{33} 
\dot{\pi}^\beta = - {\partial {\cal H}_{ADM}\over\partial\beta}, 
\end{equation} 
\begin{equation} 
\label{34} 
\dot{\pi}^\alpha = - {\partial {\cal H}_{ADM}\over\partial\alpha}, 
\end{equation} 
\begin{equation} 
\label{35} 
\dot{\pi}^\Omega = - {\partial {\cal H}_{ADM}\over\partial\Omega}. 
\end{equation} 
The first three equations are merely kinematic relations between the
parameters $\beta$, $\alpha$, $\Omega$ and their conjugate momenta
(they can be introduced in any spacetime, not necessarily a solution of
Einstein equations). The role played by these relations (\ref{31}),
(\ref{32}) is to provide a parameterization (by shift and lapse) of the
slicing and coordinatization conditions given on the initial
hypersurface to the whole spacetime. It is important to realize that
this parameterization depends on the initial-value problem and is
interwoven with the dynamics of the problem. Parameterizing the slicing
and coordinatization conditions can be achieved by different
means. However, one can convince oneself that any reasonable set of
slicing and coordinatization conditions can be translated into the
language of shift and lapse, and that such a translation provides a
proper interpretation of the conditions in terms of measurements.

Equations (\ref{31}), (\ref{32}) can be transferred to the square-root
Hamiltonian quantization procedure. One should notice, however, that
the right hand sides of these equations depend on both $\alpha$,
$\Omega$ and the true dynamic variables $\beta$ that are given, after
quantization, by distributions. The situation can be remedied if their
expectation values over the solution of the Schr\"odinger equation
(augmented by the initial-value data) are substituted in (\ref{31}),
(\ref{32}). It is clear that no contradictions can be introduced this
way (which follows from the simple count of equations and
unknowns). The relations (\ref{31}), (\ref{32}) can be thought of as
definitions of $N^i$, $N$ in quantum gravity. In practice, this
procedure can become rather complicated. In particular, the
parameterization contains functions (the expectation functions of
dynamic variables as well as of their conjugate momenta) that can be
completely determined only after the Schr\"odinger equation has been
solved. This means that parameterization of coordinatization functions
and the many-fingered time scale parameter by the shift and lapse
results in an implicit procedure in which even the parameterization
itself remains undetermined until the final solution of the entire
problem has been obtained. Nevertheless, it seems to be unavoidable if
one is to interpret the solutions of the quantum gravity equations in
terms of observations. Appropriate care should be taken, as, in
general, equation (\ref{30}) does not have to be satisfied anymore
even for the expectations, which means that, in general, the
expectation values of momenta components are not related to the
extrinsic curvature of the spacetime determined by the expectation
values of the 3-metric, shift, and lapse.  The spacetime itself is not
a solution of Einstein equations.


A brief summary of the this section is that the problems of time 
in quantum gravity apparently originate from two mutually related sources. 

The first source is a treatment of the classical gravitational field
dynamics as dynamics of the 3-metric or 3-geometry of a spacelike
slice. Both the Dirac and the square-root Hamiltonian quantization
procedures are essentially based on the original ADM\cite{ADM62}
picture of the gravitational field dynamics. In this picture, the
dynamic evolution of the gravity field manifests itself as a change
from one spacelike 3--geometry to another. In other words, the
configuration space of geometrodynamics is believed to be Wheeler's
superspace.\cite{Whe70} Such an approach does not utilize York's
analysis of the gravitational field's degrees of freedom and the
proper initial-value problem formulation.\cite{Yor72,Yor73} This is
not surprising as the foundations of quantum gravity
\cite{Per62,DeW67} were originally formulated before York completed
his investigation.

The second source of difficulties, internally related to the first
one, is the demand that the quantum gravitational dynamics should be
fully constrained, just as the dynamics of classical general
relativity.  This line of reasoning leads to the conclusion that, to
quantize the gravity field, one needs only to quantize the
constraints.

The picture of the dynamic evolution of the gravity field as a change
from one 3--geometry to another leads in classical geometrodynamics to
a peculiar situation wherein the Hamiltonian of the Hamilton-Jacobi
equation (super--Hamiltonian ${\cal H}$) do not ordinarily coincide
with the Hamiltonian participating in Hamilton evolution equations
(${\cal H}_{ADM} = N {\cal H} + N_i {\cal H}^i$). It can be argued
that other pieces of ${\cal H}_{ADM}$ are contained in the equations
of the super--momentum constraint. Nevertheless, the shift and the
lapse are thrown out of the picture, and, together with them, a
meaningful description of a time evolution of gravitational
systems. This does not cause any difficulties in the classical theory
as one can always reintroduce such a description either by retrieving
all the Hamilton evolution equations (that are compatible with the
constraints) or just four of them (the ones that provide kinematic
relations between the time derivatives of the coordinatization and the
many-fingered time functions and the conjugate momenta). One should
keep in mind, however, that these relations, ordinarily, involve all
the conjugate momenta, including those of truly dynamic variables. The
classical theory on this level treats all the variables in essentially
the same way (as functions on a spacelike slice). The situation is
entirely different for a quantum theory. Some of the variables become
the arguments of the state functional, others become the functional
parameters of the state functional. The arguments of the state
functional form the superspace. A computation of the expectation value
of an observable implies integration over the superspace, while the
functional parameters are given according to some principles that are
not a part of the dynamic picture. Ordinarily, the superspace is
formed by the dynamic variables. In Dirac's quantization all the
variables are treated on equal footing and superspace is formed by all
the components of the 3-metric. There are no functional parameters to
be used to fix coordinatization conditions or the many-fingered
time. Sometimes, as an afterthought, the superspace is reduced to the
space of 3-geometries, in which case there are three functional
parameters to fix coordinatization, but then, in order to satisfy the
super--momentum constraints, the observables are demanded to be
independent of these three functional parameters. The many-fingered
time parameter is not in this picture. If one decides to treat the
many-fingered time parameter as a part of the answer rather than of a
question, one can try to introduce it as an expectation function of one
of the 3--geometry parameters. Questions and difficulties related to
such a procedure have been outlined in the previous section.

We wish to emphasize here that this entire situation is created by the
treatment of the 3-geometry as the dynamic object and an attempt to
quantize the 3-geometry.  Mathematically, it is expressed via
imposing the commutation relations\cite{DeW67} on all the components
of the 3--metric. One should keep in mind, however, that in general
relativity the system described by the 3--metric or even the
3-geometry includes in itself the observer and his clock. In standard
quantum mechanics, or even in the quantum field theory, an observer is
external with respect to the quantum system.  The observer is
classical and has an external classical clock. A measuring device can
be coupled to a quantum system and uncoupled from it at will. There
cannot be an external observer in the description of the gravitational
field because the gravitational system is the Universe itself. An
observer cannot switch on and off the coupling of a classical
measuring device to the system. Dirac's approach essentially quantizes
the observer and his clock on equal footing with the rest of the
system (J.~A.~Wheeler would say that many-fingered time is
quantized).  Whether such a quantized observer and his clock can
function in a fashion providing an opportunity to describe the system
consistently is not clear to us. A discussion of such a possibility,
however exciting, clearly would lead us far beyond the scope of this
paper.

Another resolution to the time evolution problem in Dirac's
quantization approach would be to first solve the Wheeler-DeWitt
equation and subsequently split all six variables into (1) the two
true dynamic variables $\beta$, and (2) the remaining four embedding
variables $\alpha$ and $\Omega$. The embedding variables could then be
reserved for fixing the coordinates as well as the many-fingered time
parameter, while the true dynamic variables should be integrated over
only when the expectations of the observables are
computed. Unfortunately, such a procedure would contradict the entire
ideology behind the Dirac quantization procedure.

The ADM square-root Hamiltonian procedure of quantization avoids the
impossibility of introducing a functional time via solving the
super-momentum and super-Hamiltonian constraints on the classical
level and quantizing the resulting equations. The superspace of the
theory is thus reduced to the space of true dynamic variables with
four functional parameters left as coordinatization parameters and a
many-fingered time parameter. This procedure, however, leads to a
Schr\"odinger equation with a square-root Hamiltonian, which poses
well known technical difficulties briefly discussed above.  That no
prescription for a meaningful choice of the coordinatization and
many-fingered time variables, similar to the shift and lapse
parameterization of classical general relativity, is supplied in this
approach presents another difficulty of a conceptual nature. An
attempt to augment the procedure by such a prescription is described
above. The description of the evolution becomes implicit with a
parameterization being finalized only after the entire problem of
evolution has been solved. We wish to stress at this point that the
ADM square-root approach to gravity quantization starts, as Dirac's
approach, from a treatment of classical gravity field dynamics as the
dynamics of a 3-geometry. Exactly as in Dirac's approach, the
fully-constrained property of classical general relativity is
introduced as a key feature that is to be transferred to the quantum
theory literally. This is accomplished by turning the constraint
equations into operator equations which restrict the admissible states
of the gravitational systems.  The only difference is that the
constraints are solved before quantization in order to introduce a
Hamiltonian that, in a sense, can be interpreted as a time evolution
generator. The (functional) dimension of the superspace is reduced to
the correct dimension as determined by a proper analysis of
gravitational degrees of freedom.\cite{Yor72,Yor73} However, this
achievement comes almost as an afterthought, a fix of a deficiency
that has been created by the interpretation of the 3-geometry as the
principal dynamic object, which, in turn, has forced on the theory an
identification of the constraints with the Hamilton-Jacobi
equation. The square-root Hamiltonian participating in this equation
again differs from the Hamiltonian ${\cal H}_{ADM}$ participating in
Hamilton equations (\ref{8}), (\ref{9}) and is not related to the
Lagrangian by a relation similar to (\ref{15}). Such a switch of
Hamiltonians is possible only because the classical dynamics of general
relativity is fully constrained. In transition to the quantum theory,
the achievements of this entire procedure come at a price, creating
difficulties on both the technical and the conceptual levels.

\section{Discussion.}
\label{VII}

In this paper we have reviewed the issue of time evolution in quantum
gravity.  Both the Dirac and the ADM square-root Hamiltonian
quantization procedures create difficulties in introducing a
meaningful concept of time evolution. The difficulties in both
approaches, we conclude, stem from their common tendency to transfer
to quantum theory an interpretation of classical geometrodynamics as
an evolution of a spacelike hypersurface 3--geometry together with the
fully-constrained property of classical geometrodynamics.  These two
features essentially lead in classical geometrodynamics to an
identification of Hamilton-Jacobi equation with the constraints and
thereby removing a meaningful concept of time. This procedure does not
present a serious problem in the classical theory of gravity as time
evolution can be reinserted back.  However, it leads to two different
Hamiltonians in the theory. One of them participates in Hamilton
evolution equation, the other one is the Hamiltonian of
Hamilton-Jacobi equation.

These two features of classical general relativity, when transferred
to quantum gravity, motivate quantization of constraints (in original
form in Dirac's procedure, or resolved with respect to the momenta
conjugate to the embedding variables in the ADM square-root
Hamiltonian procedure) via imposing their operator versions on the
state functional. The resulting quantum theories naturally counter any
meaningful concept of time evolution. They lead to what Kucha\v r has
identified as the ``problems of time,'' including (1) the problem of
functional evolution, (2) the problem of multiple choice, (3) the
Hilbert-space problem of Dirac's approach and (4) the
spectral-analysis problem of the square-root Hamiltonian approach.

We have discussed the problems of time mainly for the ADM square-root
Hamiltonian quantization, as Dirac's quantization, when interpreted
literally does not admit a functional many-fingered time. Presumably,
it can be somehow reinterpreted (cf. Sec.~II), but one such avenue
leads back to the ADM square-root Hamiltonian approach.  This approach
has all the time problems in it except the Hilbert-space problem. The
way the ADM approach is ordinarily formulated leaves little room for
an analysis of the functional-evolution problem and the multiple-choice
problem. At first sight, these problems appear to be genuine. The
spectral-analysis problem seems to be unavoidable for the square-root
Hamiltonian quantization.  In addition, we conclude that the approach
suffers from the lack of a meaningful prescription for time
parameterization (although it admits functional time).  Our attempt to
reintroduce into the ADM quantization a time parameterization of
evolution by shift and lapse has lead to an implicit set of equations
with a parameterization depending on the initial values. This last feature
does not destroy the procedure but it gives one an idea of what one is to
expect if he introduces time in a way accessible to measurements.

The new idea presented in this paper is to reconsider the classical
picture of geometrodynamics via representing it as a dynamic theory of
two interacting subsystems, one being described by the dynamic
components of the 3-metric, and the other being described by the
many-fingered time variable and coordinatization variables. If one is
to keep the concept of classical time in the theory, one should
quantize only the first subsystem, while treating the second one as
classical yet generated by, and back reacting on, the first (quantum)
subsystem. In such an approach classical geometrodynamics can be
described by two sets of equations. Equations describing the evolution
of the first subsystem treat the second subsystem as an external
field. Having in mind that, eventually, we are to quantize the first
subsystem, we describe its evolution by an appropriately formulated
Hamilton-Jacobi equation. For the second system we use as field
equations super-momentum and super-Hamiltonian constraints rewritten
in appropriate variables.  The full system of equations is equivalent
to the ten Einstein equations.  However all the equations of this system are
independent.

Quantization of such a system is achieved via turning the
Hamilton-Jacobi equation into a Schr\"odinger equation. The
constraint equations describe the second subsystem as classical but
include the dynamic variables as a source. We replace them by the
expectation values of these variables over the solution of our
Schr\"odinger equation (with appropriate initial data). It is clear
that in such a procedure of quantization there is no Hilbert space
problem. The Hamiltonian of the Schr\"odinger equation does not
contain square roots which eliminates the spectral-analysis problem as
well.

The problem of functional evolution and the multiple-choice problem
obtain an interesting interpretation in our proposed approach. In our
geometrodynamic procedure of quantization they turn into a statement
that the solution of a given problem depends on the split of the
original system into its quantum and classical subsystems. In 
particular, different splits generate essentially different systems. 
In other words, quantum
geometrodynamics resolves gravitational systems finer than classical
geometrodynamics.

We have illustrated our procedure using a Bianchi~1A cosmology and an
axisymmetric Taub model.  When these models were quantized, our theory
generated quantum geometrodynamic pictures based on a post-ADM
treatment of the gravitational field dynamics and was free from the
conceptual difficulties usually associated with the Dirac and ADM
procedure of quantization. The variables describing the gravity field
in these cases have been split into the true dynamic variables and a
parameter related to Wheeler's many-fingered time. Only the dynamical
part, the underlying anisotropy, has been quantized. The Hamiltonian
participating in the Schr\"odinger equation is not a square-root
Hamiltonian. This absence of a square-root Hamiltonian is generic for
our quantum geometrodynamic procedure.  The effective ``background
spacetime'' determined by the expectation values of dynamic variables
together with the ``observers'' (related to Wheeler's many--fingered
time) allows us to pose unambiguously the questions of covariance.

The problems outlined in Sec.\ref{VI} become all but eliminated by our
quantum geometrodynamic approach.  The nontrivial part of gravity
quantization appears to shift from such conceptual problems too the
problem of (1) the choice of an appropriate model to quantize, and to
the related problem of (2) the choice of an appropriate initial
condition for the wave functional.  Both choices are crucial if one is
to attempt using quantum geometrodynamics to better comprehend the
properties of gravitational systems. It is our understanding that the
success of gravity quantization rests on such meaningful choices.
Furthermore, the choice of models should not be determined by the
structure of quantum geometrodynamics; rather, it should be determined
by observational data and our general understanding of gravitational
phenomena.

It is clear that the procedure of quantizing the dynamical part of
constrained systems described in this paper can be performed in the
general case of geometrodynamics without any complications in
principle, although it may become quite involved computationally as
compared to our simple model examples.\cite{HKM94} The procedure
differs only in two respects from the simplistic examples presented
here. The first difference arises when the 3-metric is (1)
parameterized by three coordinatization parameters, (2) the
many--fingered time parameter, and (3) the two dynamic variables; then
all four constraints should be solved with respect to the
coordinatization parameters and the scale factor. In all four
constraints the expectation values of the true dynamic variables
should be used. The second difference is caused by the functional
nature of the gravitational field dynamics in the general case.  The
operation of functional integration is involved, which might lead to
analytic difficulties. Such difficulties are not specific for our
approach as they are common for the canonical formulations of all
field theories. Quantum geometrodynamics, in particular, does not seem
to generate any specific new difficulties.  In this paper we forwarded
the beginnings of a quantization scheme consistent with York's
analysis of the gravitational degrees of freedom.  Our particular
imposition of the four constraint equations leads to a weaker theory
that in turn avoids the problems of time. 

\acknowledgements

For discussion, advice, or judgment on one or another issue taken up 
in this manuscript, we are indebted J.~Anglin, B.~Bromley, R.~Fulp, S.~Habib,
R.~Laflamme, P.~Laguna, L.~Shepley, J.~A.~Wheeler, and J.~York..

\end{document}